\documentclass[letterpaper]{article} 
\usepackage{aaai2026}  
\usepackage{times}  
\usepackage{helvet}  
\usepackage{courier}  
\usepackage[hyphens]{url}  
\usepackage{graphicx} 
\usepackage{natbib}  
\usepackage{caption} 
\usepackage{algorithm}
\usepackage{algorithmic}

\usepackage{graphicx}
\usepackage{subcaption}

\usepackage{amsmath}
\usepackage{array}
\usepackage{multirow}
\usepackage{cleveref}
\usepackage{xspace}
\usepackage{tabularx}
\newcommand{\alg}{F.A.C.U.L.\xspace}
\newcommand{\bertname}{BERT-FID\xspace}
\usepackage{newfloat}
\usepackage{listings}
\DeclareCaptionStyle{ruled}{labelfont=normalfont,labelsep=colon,strut=off} 
\floatstyle{ruled}
\newfloat{listing}{tb}{lst}{}
\floatname{listing}{Listing}
\title{F.A.C.U.L.: Language-Based Interaction with AI Companions in Gaming}
\author{
    Wenya Wei\textsuperscript{\rm 1}\equalcontrib, 
    Sipeng Yang\textsuperscript{\rm 2}\equalcontrib,
    Qixian Zhou\textsuperscript{\rm 1},
    Ruochen Liu\textsuperscript{\rm 1},
    Xuelei Zhang\textsuperscript{\rm 1}, \\
    Yifu Yuan\textsuperscript{\rm 3},
    Yan Jiang\textsuperscript{\rm 1},
    Yongle Luo\textsuperscript{\rm 1},
    Hailong Wang\textsuperscript{\rm 1},
    Tianzhou Wang\textsuperscript{\rm 1},
    Peipei Jin\textsuperscript{\rm 1},\\
    Wangtong Liu\textsuperscript{\rm 1},
    Zhou Zhao\textsuperscript{\rm 2},
    Xiaogang Jin\textsuperscript{\rm 2}, 
    Elvis S. Liu\textsuperscript{\rm 1}\thanks{Corresponding author: elvissyliu@tencent.com} 
}
\affiliations{
    \textsuperscript{\rm 1}Tencent Games, Shenzhen, China \\
    \textsuperscript{\rm 2}Zhejiang University, Hangzhou, China\\
    \textsuperscript{\rm 3}Tencent AI Lab, Shenzhen, China\\
}

\begin{document}

\maketitle

\begin{abstract}
In cooperative video games, traditional AI companions are deployed to assist players, who control them using hotkeys or command wheels to issue predefined commands such as ``attack'', ``defend'', or ``retreat''. Despite their simplicity, these methods, which lack target specificity, limit players' ability to give complex tactical instructions and hinder immersive gameplay experiences. To address this problem, we propose the FPS AI Companion who Understands Language (F.A.C.U.L.), the first real-time AI system that enables players to communicate and collaborate with AI companions using natural language. By integrating natural language processing with a confidence-based framework, F.A.C.U.L. efficiently decomposes complex commands and interprets player intent. It also employs a dynamic entity retrieval method for environmental awareness, aligning human intentions with decision-making. Unlike traditional rule-based systems, our method supports real-time language interactions, enabling players to issue complex commands such as ``clear the second floor'', ``take cover behind that tree'', or ``retreat to the river''. The system provides real-time behavioral responses and vocal feedback, ensuring seamless tactical collaboration. Using the popular FPS game \textit{Arena Breakout: Infinite} as a case study, we present comparisons demonstrating the efficacy of our approach and discuss the advantages and limitations of AI companions based on real-world user feedback.
\end{abstract}

\begin{links}
    \link{Homepage}{https://sites.google.com/view/FACUL-demo}
\end{links}

\section{Introduction}
Over the past few decades, video games have attracted billions of players, with new techniques continuously introduced to enhance their appeal. In particular, recent breakthroughs in AI have enabled the integration of advanced capabilities in games, such as those seen in Go~\cite{silver2016mastering} and real-time strategy games like StarCraft II~\cite{vinyals2019grandmaster}, where AI has reached expert-level performance. However, most AI research in gaming focuses on competition with players, while the exploration of AI companions for cooperative gameplay has been largely overlooked~\cite{gao2023towards}. In First-Person Shooter (FPS) games, players' enjoyment often depends on effective teamwork, making cooperation crucial for gameplay enhancement and immersion. When it is challenging to find human teammates online, or when players prefer to avoid teaming up with others, the development of AI companions becomes crucial. Such companions must be capable of comprehending player instructions and delivering reliable, real-time assistance.

\begin{figure}
    \centering
    \includegraphics[width=1\linewidth]{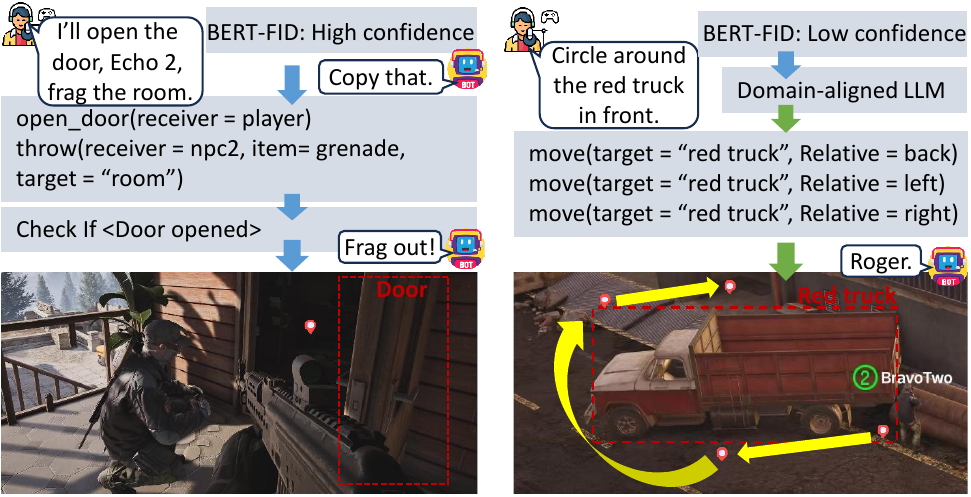}
    \caption{\alg is the first open-interaction and real-time language-operated AI companion system for commercial video games.}
    \label{fig:maincase}
\end{figure}

For AI companions, automated goal execution is relatively straightforward to implement, requiring only predefined objectives, paths, and action sequences. However, the methods for players to communicate with and command AI companions remain limited and simplistic. Existing interaction approaches, such as \textbf{hotkeys} and\textbf{ command wheels} in FPS games, allow players to issue basic commands like ``attack'' and ``retreat''. These commands are overly simplistic and lack target specificity,
which prevents players from issuing more complex tactical instructions, such as ``clear the second floor'' and ``take cover behind that tree'', hindering human-AI collaboration and reducing immersion.

Intuitively, language is the most natural and immersive form of interaction, making it highly appealing to collaborate with AI companions in games. Although recent advances in powerful large language models (LLMs) \cite{qwen2, zhang2024proagent} have enabled this capability, integrating them into real-world games for effective use presents several challenges. First, the framework must operate with minimal latency to deliver real-time feedback to players, while ensuring high concurrency and low operational costs on the server side. Second, the language model must accurately understand player commands within the specific game context, translating and decomposing commands into actions, objects, and other relevant elements. Finally, AI companions must precisely identify the objects in the game environment that correspond to the player's commands to execute them effectively.

In this paper, we propose \alg, the first language-based AI companion framework for FPS video games. \alg enables natural language interaction with players, supports tactical collaboration, and understands the game environment. The framework comprises three key components: instruction reasoning, environmental perception, and decision execution. To achieve high-throughput language instruction processing, we introduce a confidence-based method that combines a lightweight model, \bertname (BERT for Fuzzy Intent Detection), with a larger LLM. \bertname quickly decomposes commands into action–target templates and provides both reasoning results and confidence scores. High-confidence (simple) commands are executed directly, while low-confidence (complex) ones are routed to the LLM for deeper interpretation. This design reduces computational overhead for routine commands while preserving accuracy for ambiguous instructions, balancing efficiency and comprehension. Next, a multimodal entity retrieval module matches player commands to real-time game assets, ensuring robust environmental awareness. Finally, commands are translated into atomic actions and merged with environmental data to generate the AI companion’s responses, which are executed using behavior trees with real-time feedback to the player.

Using the popular FPS game \textit{Arena Breakout: Infinite} \cite{tencent2024arena} as a case study, we demonstrate the application of \alg, offering an immersive human-AI collaboration experience and supporting a wide range of open-ended commands within the game. Fig.~\ref{fig:maincase} illustrates the performance of our algorithm in task completion. A full match visualization video is available on our project page. 
We summarize the contributions of our work as follows: 
\begin{itemize}
    \item \alg, the first FPS AI companion that understands human language, offering natural language interaction and contextual environment recognition for commercial video games.
    \item \bertname, a command decomposition and confidence-based model for fuzzy intent detection, enhancing both the speed and accuracy of command processing.
    \item Developed and evaluated in a real-world FPS game, showing high reliability and strong player adoption, with a comprehensive analysis of AI companions through player feedback.
\end{itemize}

\begin{figure*}
    \centering
    \includegraphics[width=0.9\linewidth]{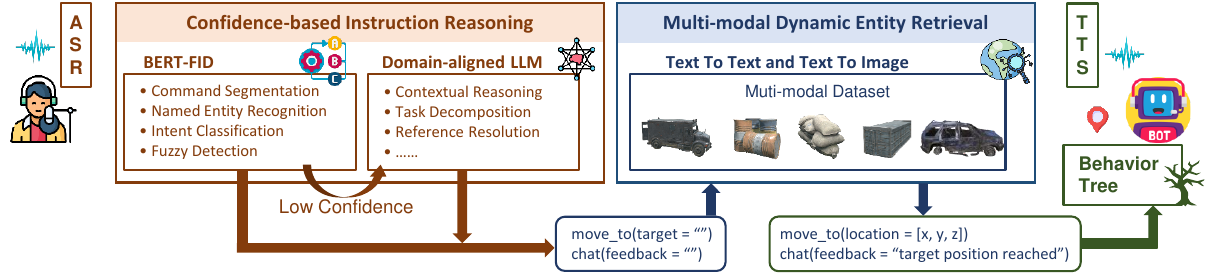}
    \caption{Overview of \alg. Players can cooperate with the agent in combat and freely speak into the microphone and get behavior response with voice feedback.}
    \label{fig:framework}
\end{figure*}

\section{Related Work}


\textbf{LLMs in gaming} have recently garnered significant attention~\cite{gallotta2024large,xu2024survey,park2023stanford}, particularly in the development of non-player characters (NPCs) and player assistance. For NPCs, LLMs are leveraged to enhance character realism and interactivity. For instance, Shao et al. \shortcite{shao2023character} propose the Character-LLM, a framework that trains LLMs to simulate specific characters, thereby enriching role-playing experiences. Similarly, the integration of LLMs in 3D games enables players to collaborate with LLM-powered NPCs to complete tasks~\cite{rao2024collaborative}. Another promising direction is the use of LLMs to generate dynamic, interactive game content. A notable application is the text-to-game engine~\cite{zhang2024text}, which transforms text inputs into dynamic RPGs, generating content such as storylines and game mechanics in real-time.

A recent commercial game, \textit{Naraka}~\cite{yongjie}, adopts a complex AI companion system, which is close to our work. However, it focuses primarily on language communication and behavioral interactions with players, while neglecting the ability to effectively understand real-time game states and identify relevant objects within these environments. This limitation restricts its ability to assist players in task completion and immersion.
In contrast, our work addresses this gap by integrating instruction reasoning, scene recognition, as well as strategic execution~\cite{meta2022human}, thereby enabling AI companions to more effectively collaborate with players and adapt to the complexity of in-game scenarios

\textbf{Instruction reasoning} refers to the ability of AI companions to understand players' language commands and intentions in gaming. Early methods~\cite{huang2015bidirectional,LiuL16} relied on traditional machine learning, which struggled with complex linguistic phenomena. The advent of pre-trained language models, particularly BERT-based methods~\cite{vaswani2017attention,kenton2019bert,shao2020bert,zhang2021unbert}, has led to significant improvements in handling downstream tasks. More recently, LLMs have demonstrated remarkable generalization capabilities. 
Models developed by \cite{Tom2020Lang} and \cite{Ajay2023Bidi} demonstrate powerful generative and comprehension abilities, enabling zero-shot learning and significantly reducing the reliance on labeled data.
Moreover, recent work~\cite{Yaniv2023Fast,LinFYBHBA0023,Chen2023acce,Chen2024data} explores collaborative paradigms between small and large models to accelerate inference and reduce deployment costs for various text-related tasks. Inspired by these advancements, we have enhanced BERT with fuzzy intent detection capabilities, leveraging the generalization strengths of LLMs to more effectively infer player intents.

\textbf{Image-text retrieval (ITR)} is a core cross-modal task that aligns visual and textual information, supporting scene recognition for AI companions. A common approach encodes image and text features separately, then aligns them via similarity computation~\cite{DeViSE,vse}. Recent advances in large-scale cross-modal pre-training have significantly improved ITR by leveraging web-scale data. For example, CLIP~\cite{clip,Yang2023chinese} uses contrastive learning on image-text pairs, while ALIGN~\cite{jia2021scaling} jointly trains encoders for efficient retrieval. In our case, we adopt a similar ITR strategy to match human instructions with game entities, further incorporating real-time attributes such as player-object distance and orientation to improve scene recognition accuracy.

\section{Overview of \alg}\label{section:overview}
Fig.~\ref{fig:framework} shows the pipeline of the proposed \alg system. In a cooperative FPS game, players interact with the \alg agent via voice input through a microphone. The speech is transcribed by an Automatic Speech Recognition (ASR) module \cite{MalikMMM21}. Then, a confidence-based instruction reasoning approach follows: \bertname processes high-confidence commands, while low-confidence (often complex or ambiguous) ones are routed to a domain-aligned LLM for deeper analysis. After extracting the player's intended action and target, the multi-modal entity retrieval module identifies relevant 3D objects in the game scene. The parsed command is then sent to the game engine, where the AI companion executes it via behavior trees and delivers real-time feedback through a text-to-speech (TTS) system~\cite{KumarKS23}. These interconnected modules form the complete \alg framework, supporting interpretability and enabling independent optimization of each component. Instruction reasoning and scene recognition are elaborated in the following sections.

\section{Confidence-Based Instruction Reasoning}\label{section:method}
We present \bertname, a fast-processing module for interpreting tactical instructions and generating confidence scores. High-confidence instructions are processed directly, while low-confidence ones are routed to the LLM for deeper reasoning. This design allows \alg to efficiently handle simple commands while reserving LLM resources for complex cases.

\subsection{\bertname: BERT with Fuzzy Intent Detection}\label{BERT with Fuzzy Intent Detection}

\begin{figure*}[h]
\centering
\includegraphics[width=0.8\linewidth]{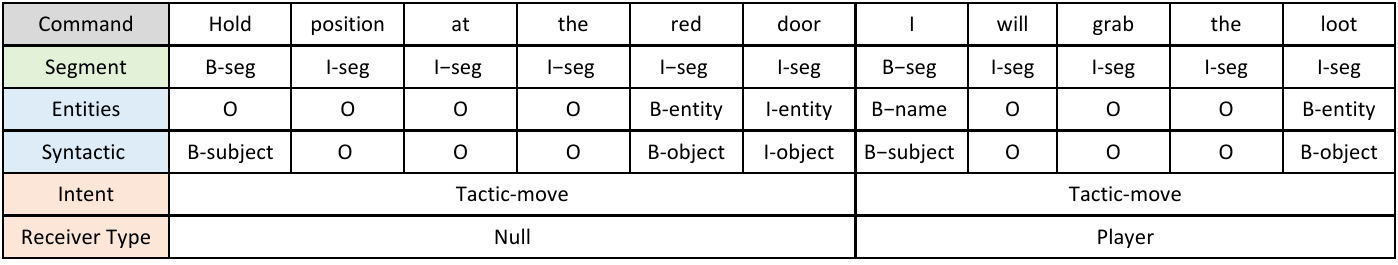}
\caption{Multi-task label structure designed for command segmentation, intent classification, and named entity recognition.}
\label{fig:bert-label}
\end{figure*}

\textbf{Multi-task Label Structure.} In gaming, player commands often need to be decomposed into a sequence of executable actions for the behavior tree, and extracting the target entities for each sub-command's intent is crucial. Based on this, we design the data label structure shown in Fig. \ref{fig:bert-label}. The segment labels in the second row are used for common segmentation tasks, where ``B-seg'' marks the beginning of a sub-command. The labels for the named entity recognition task identify entity types and syntactic categories. The intent classification labels encompass both the types of intent and their respective subjects. For example, when a player issues the command, ``Hold position at the red door, I will grab the loot'', the BERT model can parse the input, output the intent as ``move'', and identify the target location as ``the red door''.

\begin{figure}[h]
\centering
\includegraphics[width=1.0\linewidth]{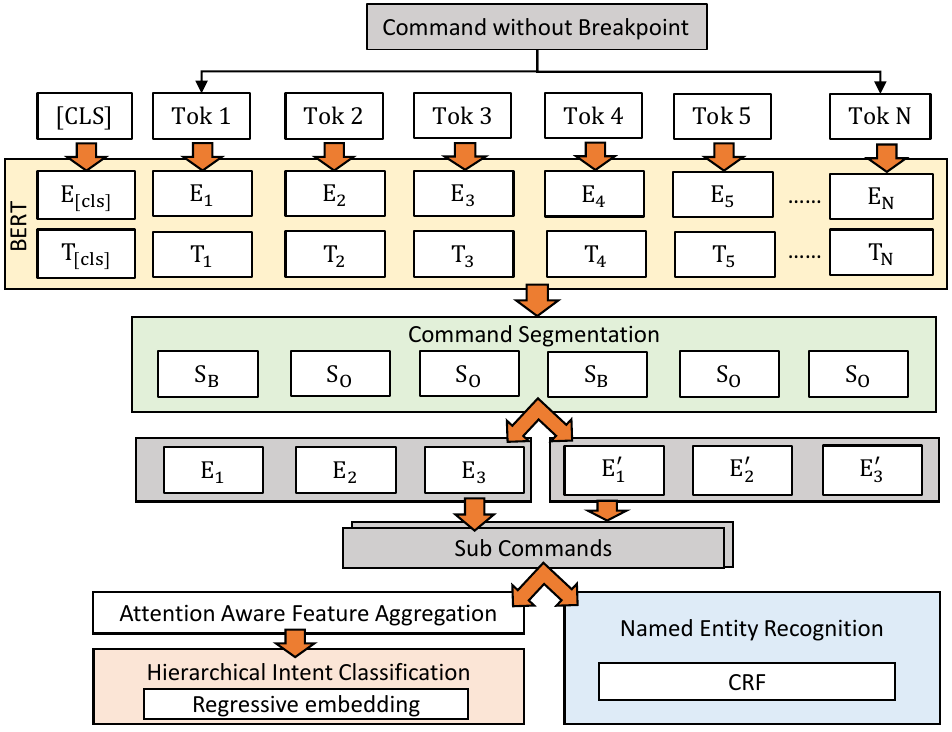}
\caption{Model Architecture of \bertname.}
\label{fig:bertmodel}
\end{figure}

\textbf{Model Architecture.} As shown in Fig. \ref{fig:bertmodel}, we use BERT~\cite{kenton2019bert} as the pre-training representation model in the feature extraction layer. For downstream tasks, we introduce command segmentation to determine whether each token starts a new sentence, thereby facilitating the precise identification of executable sub-commands. We subsequently integrate the token-level features of the segmented sub-commands through an attention mechanism to generate new features, which serve as input for subsequent hierarchical intent classification and named entity recognition. 
The recognition of named entities is treated as a sequence classification task, using Conditional Random Fields (CRF)~\cite{lafferty2001conditional} to capture contextual dependencies and accurately calculate the conditional probability distribution. In hierarchical intent classification, a regressive embedding mechanism is used across layers to improve feature extraction.

We employ this BERT base architecture for joint training across three interrelated tasks, effectively standardizing high-dimensional categories within the same data distribution. These tasks include text segmentation with a maximum length of 128 tokens, named entity recognition for six different target entities, and a hierarchical intent classification task with a three-layer intent recognition network, covering a total of 25 categories.


\textbf{Fuzzy Intent Detection in Training.}
We define the fuzzy intent as natural language commands that are either: 1) Unsupported intents for BERT model; 2) Contextual understanding; 3) Semantic-related reference resolution; 4) Instructions requiring inference. Such intents are challenging for lightweight models, as overconfident predictions often lead to incorrect execution in real-time FPS scenarios.

To enable the model to effectively leverage predicted confidence and detect fuzzy intents, we introduce an entropy optimization objective for the unsupported intent classification head in the hierarchical classifier. Specifically, given a batch of $N = N_C + N_e$ samples, where $N_c$ denotes the supported data, the cross-entropy loss is applied to compute the general loss function. For the remaining $N_e$ unsupported data, entropy maximization is used to increase the uncertainty of the probability vector. The total loss function is defined as:
\begin{equation}
\begin{aligned}
\mathcal{L} = 
& -\frac{1}{N_c}\sum_{i=1}^{N_c}\sum_{j=1}^{C} w_j y_{ij} \log p_{ij} \\
& - \lambda \frac{1}{N_e}\sum_{i=1}^{N_e} 
    \Big( \log C - \sum_{j=1}^{C} p_{ij} \log p_{ij} \Big)
\end{aligned}
\end{equation}
where $C$ is the number of categories for one head, $y_{ij}$ is the one-hot encoded label for the $j$-th category in the $i$-th sample, and $p_{ij}$ is the probability vector representing the likelihood of the $i$-th sample belonging to the $j$-th category. $w_j$ is the weight of the $j$-th category, which is determined based on the statistical distribution of intent labels $\frac{N_c}{n_i}$, where $n_i$ is the number of samples in the $j$-th category. This method increases the weight of sparse labels and reduces the weight of dense labels, which is crucial for addressing the unbalanced intent distribution in our task.

This approach allows the model to reduce confidence in any category when encountering data outside the predicted categories. In the hierarchical intent classification setup, the first-level labels determine which prediction heads are activated at the second level. For instance, if the first intent label is ``throw'', the model activates item prediction heads for grenade, smoke, food, or water. If the relevant prediction head is not activated, the model reduces confidence in any specific category by maximizing entropy regularization. The pseudocode can be found in the supplementary material.

\begin{figure}
\centering  \includegraphics[width=1\linewidth]{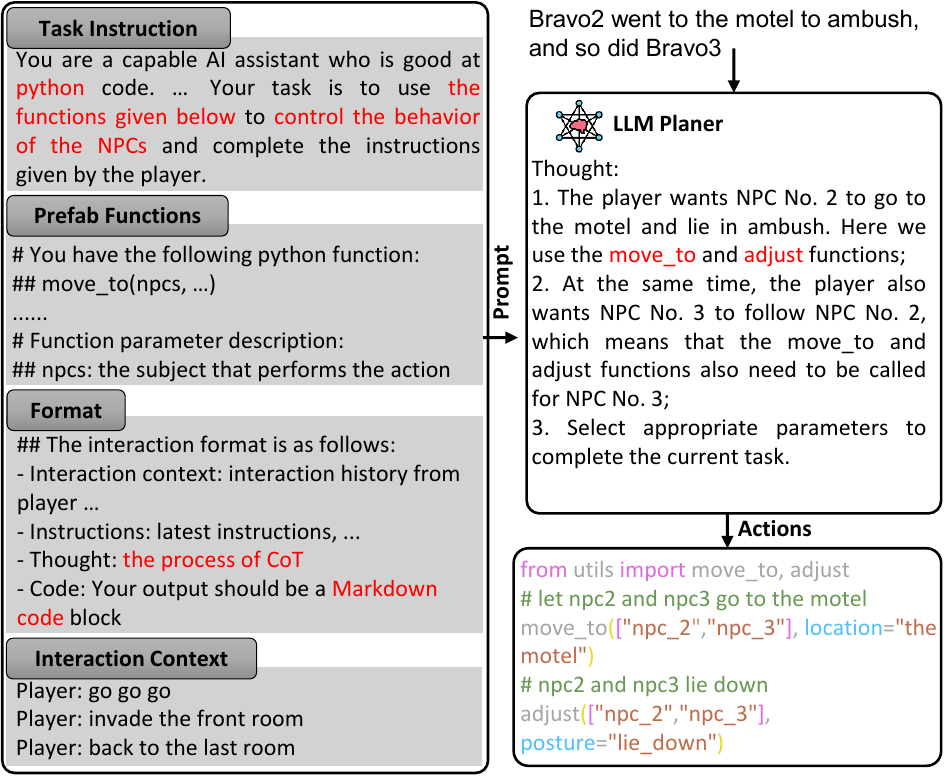}
\caption{LLM for actions planning. The LLM planner synthesizes Python code to call functions to orchestrate the actions of \alg agents, reasoning and acting to address complex tactical tasks.}
\label{fig:llm}
\end{figure}

\begin{figure*}
\centering
\includegraphics[width=1\linewidth]{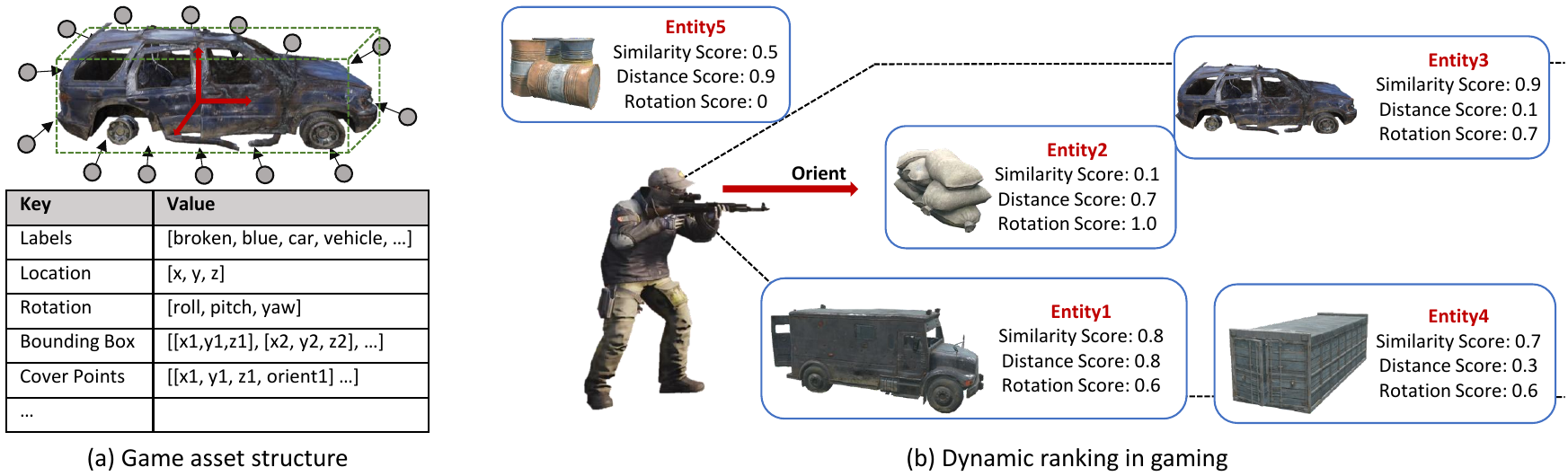}
\caption{Game asset structure and dynamic ranking. (a) Overview of 3D game assets, including labels, location, orientation, and so on. (b) Dynamic entity ranking, considering player position and command context, to identify the target entity.}
\label{fig:GameAssets}
\end{figure*}

\subsection{Domain-Aligned LLM}\label{section:llm}
To enable \alg agents to accurately execute complex or ambiguous commands that \bertname classifies with low confidence, we fine-tune a lightweight yet domain-specialized LLM using our in-house dataset for FPS games, aligning it with the preferences of real-life tactical companions. As shown in Fig.~\ref{fig:llm}, we adopt a code generation approach with LLMs to control the actions of \alg agents. Python is chosen for this task due to its compatibility with the training corpora of most existing LLMs and its flexibility in representing common code constructs, such as conditionals and loops, which JSON format lacks.

We design a series of atomic tasks (e.g., move, attack, open\_door) as Python functions, implemented via behavior trees. The LLM Planner is trained to generate Python code that invokes these functions, enabling agents to reason and execute complex tactical actions. Expected actions of FPS game agents, in response to player commands, are generated using the general-purpose LLM, Qwen2-1.5B~\cite{qwen2}, and validated by human annotators. This command-action dataset is then used to supervise the fine-tuning of the lightweight LLM.

\section{Scene Recognition}
This section details the implementation of a real-time multi-modal scene recognition system that enables fine-grained understanding of a wide range of complex objects from the player's view.
As shown in Fig.~\ref{fig:GameAssets} (a), we structure the 3D game assets dataset with tactical information, including coordinate positions, orientations, bounding boxes, and cover points. Coordinate positions are used for \alg to perform navigation, while orientation data helps \alg understand spatial relationships between game assets. Bounding boxes enable \alg to perceive the size and range of physical assets. Cover points, derived from the game's cover system, assist \alg in finding optimal positions within the FPS environment.

For entity description input, we first perform a similarity search against the embeddings of all entity labels in the scene. If the retrieved entities are insufficient, we use a fine-tuned CLIP~\cite{clip} model to search for image embeddings, generating a list of candidate entities. Real-time dynamic game data, such as the player's position and orientation, is then considered for fine-ranking the candidates. As shown in Fig.~\ref{fig:GameAssets} (b), although entity3 has the highest similarity to the command, it is too far away. Entity5 is closer but exceeds the range specified by the directional term. At last, entity1 is selected as the final result, as it ranks highest across all factors.

It is worth noting that our dynamic entity search method can handle complex hierarchical relationships. For instance, when a player issues the command: ``Go behind the red sofa on the first floor of the motel and find a cardboard box'', we first locate the primary target, ``the first floor of the motel'', then use its coordinates to search for the secondary target, ``red sofa''. Finally, the sofa is used as a reference to find the nearby ``cardboard box''.

Furthermore, since we have cover points data in the game, our entity search method can also guide \alg to seek cover at specific locations. Each object has multiple cover points, and each cover point has its corresponding orientation—only when facing the entity can it act as a cover. In this search task, we need to prioritize the orientation relationship. By assigning different weighted parameters for specific tasks, our entity search system can handle a wide variety of functions.

\section{Experiments}
To evaluate the performance of \alg, we collect real player data from the video game \textit{Arena Breakout: Infinite}. A fully deployed version of \alg is integrated into the game, and players are invited to interact with the system. This section presents two parts: a quantitative analysis comparing our method with baseline models, and a user study capturing player feedback on AI companions. 

\subsection{Quantitative Analysis}
\paragraph{Evaluation Dataset and Metrics.}\label{section:settings}
We develop a real-time test and evaluation system incorporating live player feedback. During gameplay, players issue verbal commands to AI teammates, which are transcribed into text. A thumbs-up marks successful execution, while a thumbs-down flags failures for review. These failed cases are reviewed by at least two professional annotators (more than 6 hours of FPS gaming per week), who correct the AI response. Each command is then paired with the corrected action to form the test dataset. Each game session yields 10 to 30 data points, resulting in 2,550 samples collected from 100 sessions. In our experiments, the instruction reasoning modules are executed on two NVIDIA GeForce RTX 3080 GPUs, while entity retrieval tasks run on two NVIDIA Tesla T4 GPUs. Detailed experimental settings and reference code are available on our project page. The evaluation metrics include:
\begin{itemize}
    \item \textbf{Accuracy}: Measures the proportion of correct actions executed by the AI in response to player commands.
    \item \textbf{Parameters}: Refers to the total number of parameters in the ACLI model, indicating the model’s complexity and computational demands.
    \item \textbf{Precision, Recall, and F1 Score}: Evaluate the performance of the entity retrieval models.
    \item \textbf{Queries Per Second (QPS)}: Measures the system's ability to process player commands in real-time, indicating how well it can handle high-frequency interactions. 
    \item \textbf{Latency}: Tracks the delay between a player’s input and the AI’s response.
\end{itemize}

\paragraph{Comparison and Results.}
The \alg method includes two main components: instruction reasoning and entity retrieval. We first evaluate each component separately against corresponding baselines, followed by an assessment of the integrated system.

For the \textit{\textbf{instruction reasoning}} component, textual commands are fed into the \bertname model to generate an action sequence. If the confidence score falls below a predefined threshold, the command is routed to a domain-aligned LLM to re-infer the action. We compare the performance of using BERT alone, LLM alone, and combinations of both. The evaluated models include the original BERT, \bertname, Qwen2-7B, Qwen2-1.5B, and our fine-tuned domain-aligned LLM* (based on Qwen2-1.5B). We also assess the performance of combining the original BERT with the domain-aligned LLM and \bertname with the domain-aligned LLM.

\begin{table}
\small
\renewcommand{\arraystretch}{1.16}
  \begin{tabular}{>{\centering}p{26mm}|
    >{\centering}p{10.5mm}|
    >{\centering}p{12.5mm}|
    >{\centering}p{7mm}|c}
    \hline\hline
    \multirow{2}{*}{Method} & \multirow{2}{*}{Accuracy}  & \multirow{2}{*}{Parameters} & \multicolumn{2}{c}{Decision count} \\\cline{4-5}
    ~ & ~  & ~ & BERT & LLM\\\hline
    BERT & 0.714  &  \multirow{2}{*}{\textbf{104M}} &  \multirow{2}{*}{2550} & \multirow{2}{*}{0} \\ \cline{1-2}
    \bertname & 0.749 & ~ & ~ & ~  \\\hline
    Qwen2-7B & 0.807& 7B &  \multirow{3}{*}{0} & \multirow{3}{*}{2550} \\\cline{1-3}
    Qwen2-1.5B &0.642 & \multirow{2}{*}{1.5B}  & ~ & ~\\ \cline{1-2} 
    LLM* & \textbf{0.908} & ~  & ~ & ~\\\hline
    BERT \& LLM* & 0.831 & \multirow{2}{*}{1.6B}~ &2321  & 229 \\\cline{1-2}\cline{4-5}
    BERT-FID \& LLM* & 0.886  & ~ & {2148}  & {402} \\\hline\hline
  \end{tabular}
\footnotesize{Note: ``Decision count'' indicates the number of final decisions selected from either BERT or LLM. ``LLM*'' refers to our domain-aligned LLM (Qwen2-1.5B)}
\renewcommand{\arraystretch}{1}
\caption{The comparison of various methods in about instruction reasoning demonstrates that our approach, ``BERT-FID with fine-tuned LLM'', can route more samples to the LLM and effectively achieve higher accuracy than the pure BERT-based method.}
\label{table-exp-bert}
\end{table}

\begin{table}
\small
\centering\renewcommand{\arraystretch}{1.16}
\begin{tabular}{>{\centering}p{12.5mm}|p{32mm}|c|c}
\hline\hline
Module  & Method & QPS  & Latency (ms) \\ \hline
    ~ & BERT-based method & 1672 & 412 \\ \cline{2-4}
    Instruction & Qwen2-7B & 46  &423   \\ \cline{2-4}
    reasoning & LLM* & 245 & 392 \\\cline{2-4}
    ~ & BERT-FID with LLM* & 813 &470  \\ \hline
    Dynamic& CLIP-based method & 194 &\textbf{355}  \\ \cline{2-4}
    entity & Text similarity & \textbf{2128} & 364 \\ \cline{2-4}
    retrieval & Text similarity with CLIP & 1057  &363 \\ \hline\hline
\end{tabular}
\renewcommand{\arraystretch}{1}
\caption{Performance testing shows our method achieves about 4$\times$ higher inference concurrency than LLM-based methods and five-fold higher than CLIP-based methods.}
\label{table-exp-qps}
\end{table}

Table~\ref{table-exp-bert} presents the accuracy of different instruction reasoning methods. Our enhanced \bertname outperforms the original BERT (0.749 vs. 0.714). Among LLMs, our domain-aligned model significantly improves over the Qwen2-1.5B baseline by 26.6\%, and even surpasses the larger Qwen2-7B by 9.9\%. Table~\ref{table-exp-qps} shows the runtime performance. While BERT-based models achieve the highest speed (1,672 QPS), Qwen2-7B and our domain-aligned LLM reach only 46 and 245 QPS, respectively. To balance speed and accuracy, we combine BERT-FID with the domain-aligned LLM, achieving 813 QPS with only a 2.2\% accuracy drop compared to the LLM-only setup. In contrast, combining the original BERT with LLM leads to a 7.7\% accuracy drop, further validating the effectiveness of our proposed instruction reasoning strategy.

For evaluating the \textit{\textbf{entity retrieval}} component, we compare our method with several baselines: general CLIP, fine-tuned CLIP, a text similarity method, and our integrated approach combining text similarity with fine-tuned CLIP.

\begin{table}
\small
\renewcommand{\arraystretch}{1.16}
\begin{tabular}{c|>{\centering}p{28mm}|c|c|c}
\hline\hline
Range & Method  & Precision & Recall & F1 \\ \hline
\multirow{5}{*}{\rotatebox{0}{Top 5}} & General CLIP &0.459&0.400&0.429  \\
\cline{2-5}
~&Fine-tuned CLIP &0.817 & 0.763 & 0.790 \\
\cline{2-5}
~&Text similarity &  0.904 &0.873&0.888  \\
\cline{2-5}
~&{Text similarity with}& \multirow{2}{*}{\textbf{0.926}}&\multirow{2}{*}{\textbf{0.890}}&\multirow{2}{*}{\textbf{0.908}} \\
   &{fine-tuned CLIP} & & & \\
\hline
\multirow{5}{*}{\rotatebox{0}{Top 10}} & General CLIP &0.578 &0.521 &0.549 \\
\cline{2-5}
~&Fine-tuned CLIP &0.875 & 0.842 & 0.859  \\
\cline{2-5}
~&Text similarity &  0.930 &0.913&0.921\\
\cline{2-5}
~&{Text similarity with}& \multirow{2}{*}{\textbf{0.953}}&\multirow{2}{*}{\textbf{0.932}}&\multirow{2}{*}{\textbf{0.943}} \\
    &{fine-tuned CLIP} & & & \\
\hline
\multirow{5}{*}{\rotatebox{0}{Top 20}} & General CLIP &0.688&0.636&0.662\\
\cline{2-5}
~&Fine-tuned CLIP &0.923 & 0.904 & 0.913  \\
\cline{2-5}
~&Text similarity &  0.946 &0.936&0.941   \\
\cline{2-5}
~&{Text similarity with}& \multirow{2}{*}{\textbf{0.966}}&\multirow{2}{*}{\textbf{0.954}}&\multirow{2}{*}{\textbf{0.960}} \\
    &{fine-tuned CLIP} & & & \\
\hline\hline
\end{tabular}
\renewcommand{\arraystretch}{1}
\caption{Results on various retrieval ranges and various methods in environment recognition. By combining text matching with the fine-tuned CLIP model, our method can improve the accuracy across several retrieval ranges.}
\label{table-exp-env}
\end{table}

\begin{figure*}[t]
    \centering
    \includegraphics[width=0.9\linewidth]{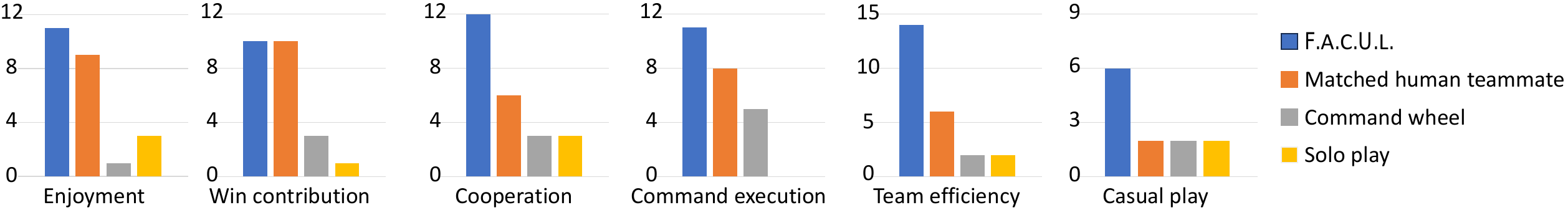}
    \caption{User study results on player preferences for four game modes across six key aspects.}
    \label{fig:us}
\end{figure*}

Table \ref{table-exp-env} displays the results of these methods across retrieval ranges of top 5, 10, and 20\footnote{We use the entity retrieval method to give a list of potential entities and then select the most suitable game asset from this list based on real-time in-game data such as the player's orientation. Therefore, this method necessitates $n \geq 5$, ensuring a more comprehensive selection process than merely opting for the top 1 result.}. Across all ranges, CLIP fine-tuned with our in-house asset label dataset significantly outperforms the general CLIP. The text similarity method also achieves good retrieval results. Furthermore, our combined approach consistently delivers the best outcomes, with F1 scores exceeding 0.9. Although this method results in a lower QPS compared to using only the text similarity method (as shown in Table \ref{table-exp-bert}), the operational speed remains within an acceptable range. As enhancing retrieval accuracy is our primary goal, this method effectively achieves it.


Finally, we evaluate the \textbf{\textit{complete \alg model}}. For comparison, we also test a fast version using \bertname with text similarity, and a slower version using LLM with CLIP.

Table~\ref{table-exp-main} indicates that by integrating four model inferences, \alg achieves an accuracy of 87.2\% and a QPS of 916, with an average response time of approximately 613 ms, providing a high-precision, cost-effective solution for commercial games. All three approaches have similar latency ($\sim$600 ms). The BERT-FID and text similarity version reaches the highest speed (2050 QPS) but has the lowest accuracy—about 15\% lower than the other two. In contrast, our \alg model reduces accuracy by only 2\% compared to the full-large-model version, while achieving a 4.5× gain in QPS. These results demonstrate that our approach effectively balances accuracy and speed, making it practical for real-world deployment.

\begin{table}
\small
\centering\renewcommand{\arraystretch}{1.16}
\begin{tabular}{c|c|c|c}
\hline\hline
Method&  Accuracy& QPS & Latency \\ \hline
BERT-FID and text similarity &73.9\%  & \textbf{2050} & \textbf{599} \\\hline
LLM* and fine-tuned CLIP &\textbf{89.2\%} & 210 & 614  \\\hline
{\alg (Our full method)} & 87.2 \%  & 916 & 613  \\\hline\hline
\end{tabular}
\renewcommand{\arraystretch}{1}
\caption{Evaluation of intent and environment recognition.}
\label{table-exp-main}
\end{table}

\subsection{User Studies}\label{section:user study}

\paragraph{Study Design.}
To assess the real-world effectiveness and user reception of our AI companions, we conduct a user study within \textit{Arena Breakout: Infinite}, comparing four collaboration modes: 1) our proposed \alg system, 2) random human teammate matching, 3) wheel-based teammate control, and 4) solo play. Participants are selected based on frequent gameplay and prior experience with \alg, ensuring familiarity with all modes. A total of 24 regular players participated in this preliminary study. The study evaluates player preferences across six key aspects: 1) most enjoyable mode, 2) mode contributing most to winning, 3) best for teammate interaction, 4) most accurate in command execution, 5) most efficient in collaboration, and 6) most preferred for casual play. 


\paragraph{Study Results.}
Fig.~\ref{fig:us} illustrates user preferences across the six key aspects. The results show that \alg and random human teammate matching are the most preferred modes for overall enjoyment, with \alg particularly favored for its superior performance in teammate interaction, command execution accuracy, and collaboration efficiency. Participants report that \alg’s integration of natural language and environmental context provides a more immersive and intuitive experience. In contrast, the command wheel mode consistently ranks lowest in most aspects, including enjoyment, interaction, and collaboration efficiency, indicating it is less appealing to players seeking a dynamic and interactive experience. Regarding perceived assistance in winning, both \alg and random human teammate matching are seen as the most helpful, offering better synergy and coordination. Solo play and the command wheel mode are regarded as less effective in this regard. For casual play, participants overwhelmingly favor \alg, highlighting its suitability for more relaxed and enjoyable gaming experiences.

In summary, the results demonstrate that \alg excels in effective interaction, command execution accuracy, and collaboration efficiency, making it a preferred mode among players. These advantages contribute to its popularity, positioning \alg as a highly practical option that enhances player satisfaction in both competitive and casual gaming experiences.


\section{Conclusion and Future Work}\label{conclution}
This paper presents \alg, a cost-effective and high-precision solution for AI companions with advanced natural language understanding and visual perception. It demonstrates how AI companions can assist players in achieving their goals and enhance the gaming experience, offering a practical solution for commercial cooperative games. We propose BERT-FID, which accounts for the uncertainty of complex and ambiguous inputs, enabling effective collaborative inference with LLMs in terms of both accuracy and speed. For scene recognition, we introduce a multi-modal dynamic entity retrieval system that aligns human intentions with decision-making elements in 3D game environments. A real-world study shows that \alg understands 87.2\% of natural language commands and identifies dynamic entities, resulting in a 77\% reduction in inference deployment costs compared to baselines.

\alg provides an efficient framework for AI teammates in FPS games by leveraging the built-in game engine, with a clear applicable scope tied to its latency: it excels in real-time collaborative FPS scenarios (100–800ms latency acceptable).  It also has potential in RPGs and MOBAs with minimal adaptations (e.g., adjusted intent templates, game asset datasets), requiring only BERT datasets, LLM prompts, and game asset datasets from developers. Future work could focus on integrating audio signals to enhance voice processing and reduce reasoning latency. Furthermore, we plan to incorporate distinct personality traits for AI companions, allowing for more engaging player interactions.

\section{Acknowledgments}
Xiaogang Jin was supported by the National Natural Science Foundation of China (Grant No. 62472373).

\section{Ethical Statement}
Informed consent was obtained from all participants prior to their involvement in the user study. Participants were fully informed about the study's purpose, procedures involved, and any potential risks and benefits. All data collected were anonymized to maintain confidentiality, with access restricted to the research team. The behavior of NPCs within games is dictated by developers' intentions and undergoes a thorough review to ensure compliance with social and ethical standards.

We are committed to ensuring our research benefits users while minimizing potential harm. AI companions are designed to enhance gaming experiences and engage players positively. Extensive testing, ongoing player feedback through surveys, community discussions, and game analyses are conducted to ensure that AI behaviors within games do not promote real-world violence or aggressive behaviors. Additionally, any potential psychological impacts on players are closely monitored and addressed.
\bibliography{aaai2026}

\clearpage
\section{Appendix}

\subsection{Background} 

\begin{figure}[h]
    \centering
    \includegraphics[width=\linewidth]{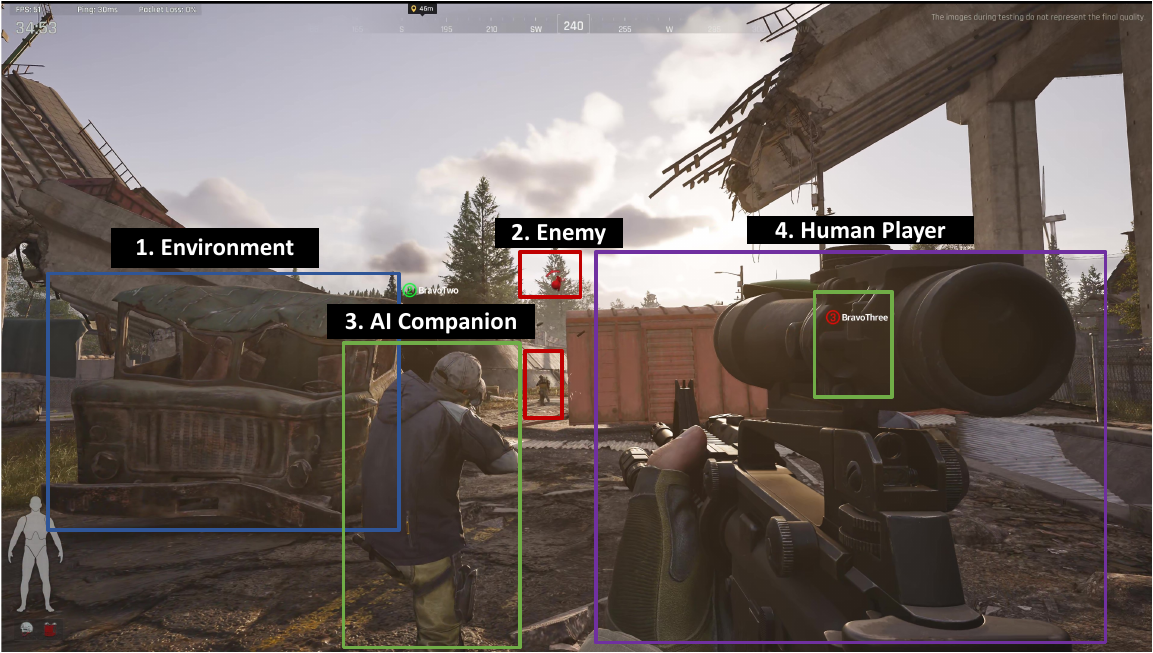}
    \caption{Game Introduction to \textit{Arena Breakout: Infinite}.}
    \label{fig:background}
\end{figure}

\textit{\textbf{Arena Breakout: Infinite}} is an immersive and highly tactical FPS game developed and published by Tencent Games. Fig.~\ref{fig:background} shows a brief introduction to the game, which illustrates important environmental information that AI teammates need to comprehend during gameplay. Players may adopt different strategies—some engage in direct combat to loot gear from enemies, while others prioritize stealth, exploration, and task completion before extraction. However, a lack of tactical consensus among randomly matched human teammates often disrupts cooperation, presenting an opportunity to introduce AI companions to improve player experiences. 

\subsection{Methods Details}

\textbf{\bertname: BERT with Fuzzy Intent Detection}

Algorithm~\ref{alg:compute_loss_entropy} presents the pseudocode for our proposed BERT-FID training algorithm, which aims to optimize the loss function for intent classification, particularly focusing on handling unsupported intent samples.

Initially, the algorithm processes a batch of samples, comprising both supported and unsupported samples. For each category, it initializes weights based on intent label statistics. Subsequently, it calculates the cross-entropy loss for supported data by updating the loss using the weights and predicted probabilities for each category. 

Following this, the algorithm computes the entropy loss for unsupported data by calculating the entropy of each sample's predicted probability vector and updating the loss accordingly. Finally, the cross-entropy loss and the scaled entropy loss are combined to yield the final loss value. 

\begin{figure*}[h]
    \centering  \includegraphics[width=1.0\linewidth]{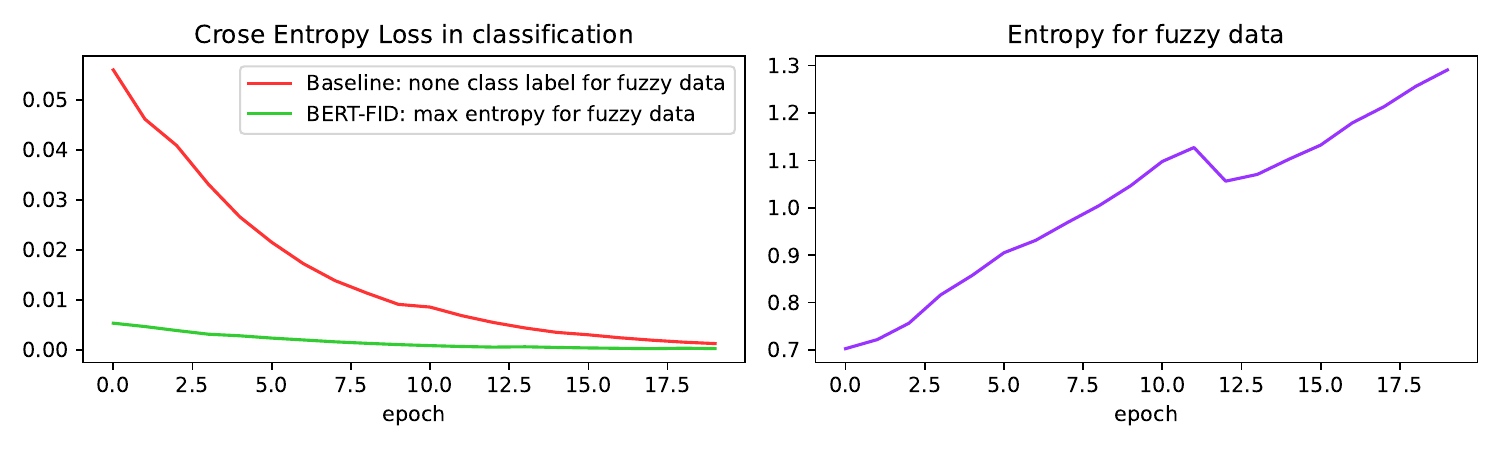}
    \caption{The train process of BERT-FID}
    \label{fig:bert train}
\end{figure*}

This loss is used to train the BERT-FID model for hierarchical intent classification tasks. Through this approach, the algorithm effectively manages unsupported intent samples, enhancing the robustness and accuracy of the classification model. We present the training process of \bertname in Fig.~\ref{fig:bert train}, which demonstrates that our training approach reduces the burden on the classifier while improving its capacity to manage uncertainty in fuzzy data.








    



\begin{algorithm}[h]
\caption{Entropy Optimization for Unsupported Intent Classification}
\label{alg:compute_loss_entropy}
\begin{algorithmic}
  \REQUIRE Batch of samples $X = \{x_1, x_2, \ldots, x_N\}$, where $N = N_C + N_e$
  \ENSURE Optimized loss $\mathcal{L}$ for intent classification
  
  \STATE $N_c \gets$ Number of supported samples
  \STATE $N_e \gets$ Number of unsupported samples
  \STATE Initialize weights $w_j$ for each category based on intent label statistics
  
  \FOR{each category $j$}
    \STATE Calculate $w_j$ as the ratio of supported samples to the number of samples in category $j$
  \ENDFOR
  
  \STATE \textbf{Calculate Cross-Entropy Loss for Supported Data:}
  \STATE Initialize $\mathcal{L}_{ce} \gets 0$
  
  \FOR{each supported sample $i$}
    \FOR{each category $j$}
      \STATE Update $\mathcal{L}_{ce}$ using the weight $w_j$ and the predicted probability $p_{ij}$
    \ENDFOR
  \ENDFOR
  
  \STATE \textbf{Calculate Entropy Loss for Unsupported Data:}
  \STATE Initialize $\mathcal{L}_{ent} \gets 0$
  
  \FOR{each unsupported sample $i$}
    \STATE Calculate the entropy based on the predicted probability vector $p_i$
    \STATE Update $\mathcal{L}_{ent}$ with the calculated entropy
  \ENDFOR
  
  \STATE \textbf{Combine Losses:}
  \STATE $\mathcal{L} \gets \mathcal{L}_{ce} + \mathcal{L}_{ent}$ as the sum of cross-entropy loss and scaled entropy loss\; 
  Train the hierarchical intent classification task for BERT-FID\;

  \RETURN $\mathcal{L}$
\end{algorithmic}
\end{algorithm}

\begin{figure*}[h]
    \centering  \includegraphics[width=\linewidth]{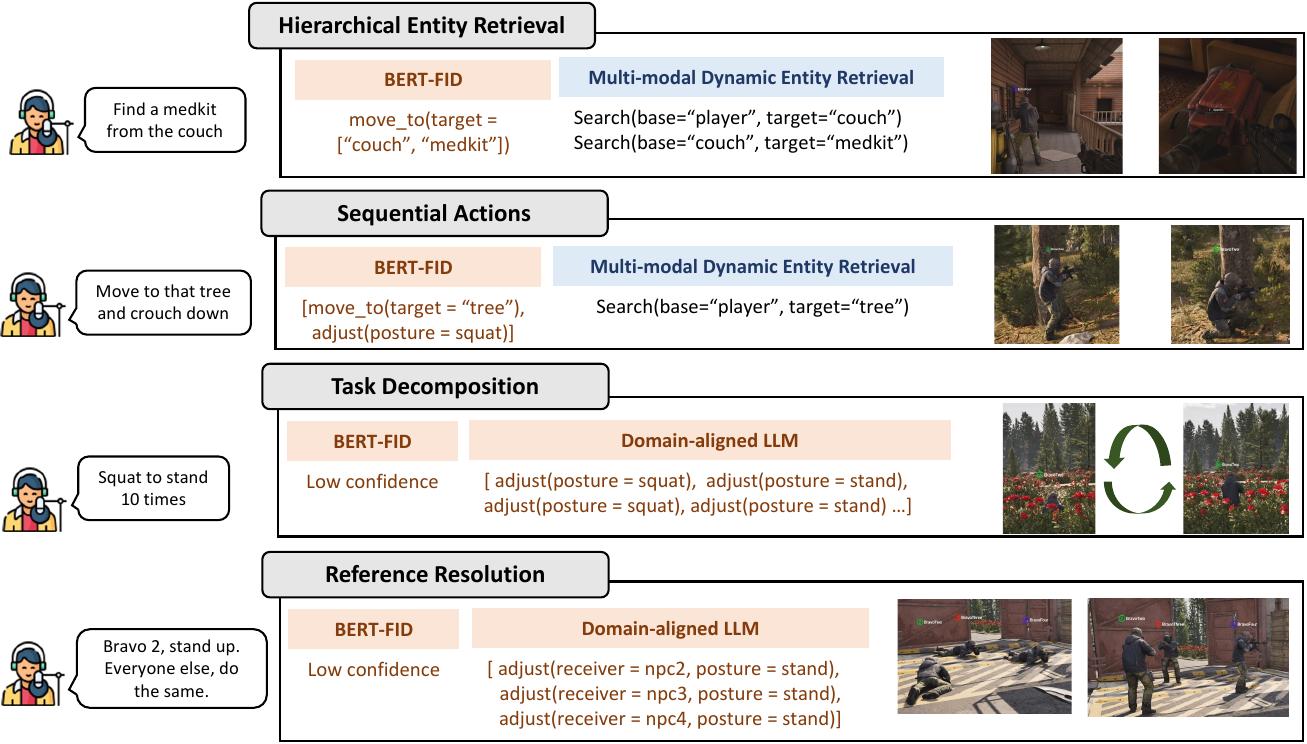}
    \caption{Four typical cases illustrate how the \alg system accomplishes complex tasks, with the BERT-FID, LLM, and entity retrieval modules collaborating to provide real-time behavioral responses in the game.}
    \label{fig:typical cases}
\end{figure*}

\textbf{Multi-modal Instructional Reasoning} In Algorithm~\ref{alg:DS_inference}, we present the integration method for Instructional Reasoning and Multi-modal Dynamic Search. 

The algorithm begins by receiving a query from a human player and uses the BERT-FID model to determine the intent and target, while also calculating the overall confidence level. If this confidence is below a certain threshold, the algorithm employs a generative language model (LLM) to generate the intent and target. The target actions include elements such as the receiver, decision, posture type, and move type.

If a target is identified, the algorithm retrieves the top-k data pairs from the asset dataset and calculates their confidence levels. If the confidence is below the threshold, the CLIP model is used for more precise text-to-image retrieval matching. The retrieved assets are then reordered based on dynamic game information, such as position, orientation, bounding box, and cover points, and the specific target location is set within the behavior tree.

This approach allows the algorithm to effectively handle multi-modal information, achieving rapid and efficient instruction parsing. By leveraging both textual and visual data, the system enhances its decision-making capabilities, ensuring accurate and contextually relevant actions within the game environment.

These are specific values that indicate the percentage of requests that are completed within a certain time frame. For example, the 90th percentile response time means that 90\% of the requests were completed in that time or less.


\begin{algorithm}[htbp]
\caption{Instructional Reasoning and Multi-modal Dynamic Search}
\label{alg:DS_inference}
\begin{algorithmic}[1]
  \REQUIRE Fast process model BERT-FID $F_{bert}$, generative language model $F_{llm}$, threshold $\delta_1$, text similarity calculator $F_{text}$, text to image similarity calculator $F_{image}$, threshold $\delta_2$
  \REQUIRE Query from human player $x$ (string format)
  \ENSURE Executable action $A$ encoding sequence for game engine

  \STATE Get intent and target by $F_{bert}$, and compute the overall confidence $C_{bert}$
  \IF{$C_{bert} < \delta_1$}
    \STATE Get intent and target by the generation of $F_{llm}$
  \ENDIF

  \STATE Set action $A$ to include receiver, decision, posture type, move type, etc.

  \IF{has target}
    \STATE Retrieve the top-k data pairs from the asset dataset using $F_{text}$
    \STATE Compute the confidence $C_{text}$
    \IF{$C_{text} < \delta_2$}
      \STATE Retrieve the top-k data pairs from the asset dataset using $F_{image}$
    \ENDIF
  
    \STATE Reorder the retrieved assets based on the dynamic game information, including position, orientation, bounding box, cover point, etc.
    \STATE Set the specific target location to action $A$
  \ENDIF
\end{algorithmic}
\end{algorithm}


\subsection{Case Study}

In order to conduct a better qualitative analysis of our system, we examine the capabilities of the \alg system in handling hierarchical entity retrieval, sequential actions, task decomposition, and reference resolution, all within a complex interactive environment. The \alg system leverages BERT-FID for feature identification and a domain-aligned Large Language Model (LLM) for command reasoning and contextual understanding, allowing it to effectively process a range of tactical instructions.

\textbf{Hierarchical Entity Retrieval.} In the first scenario, \alg needs to retrieve a medkit from a couch. It uses BERT-FID to identify the target (move\_to(target = [``couch'', ``medkit''])) and employs multimodal dynamic entity retrieval to find both the couch and the medkit in the environment. This approach allows the system to effectively navigate and interact with multiple dynamic objects in real time, improving its capability to perform complex retrieval tasks.

\textbf{Sequential Instruction.} When instructed to ``Move to that tree and crouch down," \alg efficiently combines both the movement (move\_to(target = ``tree")) and posture adjustment (adjust(posture = squat)) into a single sequential action. The system's entity retrieval capabilities allow it to dynamically adjust to environmental changes, ensuring smooth execution of multi-step commands.

\textbf{Task Decomposition.} \alg also excels at breaking down complex commands into manageable tasks. For instance, when asked to ``Squat to stand 10 times," the system identifies the need for repeated posture changes and executes this action iteratively. In cases where the confidence level of BERT-FID is low, the domain-aligned LLM steps in to ensure precise task decomposition, guiding the agent through alternating postures to complete the task.

\textbf{Reference Resolution.} In the final scenario, \alg processes an ambiguous command, ``Bravo 2, stand up. Everyone else, do the same." Here, the domain-aligned LLM takes over to assign the correct actions to specific entities (adjust(receiver = npc2, posture = stand) for Bravo 2 and subsequently applies the same action to all other NPCs). This showcases \alg’s ability to handle reference resolution in multi-agent settings.


Overall, this case study demonstrates \alg's proficiency in navigating hierarchical entity retrieval, performing sequential actions, decomposing complex tasks, and resolving ambiguous references, which are essential for enhancing the interactive depth of gaming environments.

Additional example interactions and their analysis is available in the Fig. \ref{fig:typical cases} and Table. \ref{tab:addlabel}.

\subsection{Failure Cases}

Despite its strengths, \alg has its limitations. The following examples illustrate scenarios where the system may struggle to accurately interpret or execute commands, potentially leading to failures:

\textbf{Unresolvable Commands.} Due to the limitations of the underlying implementation logic, the response actions in the behavior tree are finite. For example, if a player asks the AI to help with a countdown, but the countdown function is not pre-defined in the behavior tree interface, the AI can only refuse the command.

\textbf{Unlabeled Entity Information in Scenarios.} Commands that include slang, idioms, or unconventional phrasing can present challenges. For example, a user saying ``Go to the richest building" may not be correctly interpreted; the model may struggle to grasp the relevant context and fail to understand the player's expression. However, this can largely be addressed through ongoing player testing and iterative improvements to accommodate established gaming jargon.

\textbf{Spatial Awareness in Architectural Contexts.} Current text-image retrieval models struggle with the three-dimensional perception of space. For example, if a player’s target is ``the second window from the left on the second floor," we cannot effectively retrieve this based on spatial relationships. We hope to establish a deeper connection between 3D models and text matching.

These failure cases highlight the critical opportunities and challenges for AI teammate systems. By identifying and analyzing these issues, we can focus efforts on enhancing \alg's performance and increasing user satisfaction in future iterations.

\begin{table*}
\caption{Additional successful cases in game.}
\small
\centering
\begin{tabularx}{\textwidth}{|>{\raggedright\arraybackslash}p{0.21\textwidth}|>
{\raggedright\arraybackslash}p{0.73\textwidth}|}
\cline{1-2}
\textbf{Interaction Type} & \textbf{Human-AI Interaction} 
\\ \cline{1-2}
\multirow{6}{*}{Simple Actions} & 
Player: Stay on me.\newline
AI: Copy that. \newline
AI: \textless follow(player) \textgreater \\
\cline{2-2}
&
Player: Check the blue box on the left. \newline
AI: Roger that.\newline
AI: \textless move\_to(``left, blue box'')\textgreater \\ 
\cline{2-2}
&
Player: Give me a bottle of water. \newline
AI: Copy.\newline
AI: \textless move\_to(player), item(``water'')\textgreater \\ 
\cline{2-2}
&
Player: smoke the exfile. \newline
AI: Roger that.\newline
AI: \textless move\_to(``escape point''), item(``smoke'')\textgreater \\ 
\cline{2-2}
\cline{1-2}
\multirow{6}{*}{Sequential Actions} &
Player: Move to the green car, and then come back.\newline
AI: Copy that.\newline
AI: \textless move\_to(green car), follow(player)\textgreater \\ 
\cline{2-2}
&
Player: Place a medkit at the nearby cover location.\newline
AI: Roger that.\newline
AI: \textless move\_to(coverpoint), item(medkit)\textgreater \\ 
\cline{2-2}
& 
Player: Hold your fire, keep going! \newline
AI: Copy that.  \newline
AI: \textless adjust(not fire), move\_to(coverpoint)\textgreater\\
\cline{2-2}
&
Player: Cover fire in your position, I flank them.\newline
AI: Engaging.\newline
AI: \textless attack(enemy)\textgreater\\
\cline{1-2}
\multirow{6}{*}{Commands Reasoning} &
Player: I'm hurt.\newline
AI: Here is a medkit for you. \newline
AI: \textless move\_to(player), item(medkit)\textgreater  \\ 
\cline{2-2}
&
Player: I'm thirsty.\newline 
AI: Here is a bottle of water for you.\newline 
AI: \textless move\_to(player), item(water) \textgreater \\
\cline{2-2}
& Player: I'm low on health.\newline
AI: A medkit has been placed for you. \newline
AI: \textless item(medkit) \textgreater \\
\cline{2-2}
& Player: I'm surrounded!\newline
AI: \textless protect(player) \textgreater \\
\cline{2-2}
& Player: Something's not right about this door.\newline
AI: I'll check out. \newline
AI: \textless move\_to(door)\textgreater \\
\cline{2-2}
&Player: Bravo2,keep your position. Everyone else, come over here.
\newline
AI-2: \textless stop \textgreater \newline
AI-3: \textless move\_to(player) \textgreater  \newline
AI-4: \textless move\_to(player) \textgreater \\
\cline{2-2}
& Player: Go to the motel.\newline
AI-2: \textless move\_to(motel)\textgreater\newline
Player: Come Back.\newline
AI-2: \textless move\_to(player)\textgreater\newline
Player: Go to the place I just asked you to go.\newline
AI-2: \textless move\_to(``motel'')
\textgreater \\
\cline{1-2}
\multirow{6}{*}{Information Feedback} &
Player: What does the red smoke represent?\newline
AI: The red smoke is an evacuation point. You can escape from here. \\
\cline{2-2}
&
Player: How to get out? \newline
AI: Find the extraction point, and survive. That's the only way you're getting out of here.\\
\cline{2-2}
& 
Player: Loot the body for me.\newline
AI: Negative. It's all yours, I'll cover you. \\ 
\cline{2-2}
&
Player: Give me a key to the main guest room. \newline
AI: Negative. That key costs 900K. I can't afford it.\\
&
Player: Bravo2, attack Bravo 3 \newline
AI: Negative. No friendly fire!\\
\cline{1-2}
\end{tabularx}%
\label{tab:addlabel}%
\end{table*}

\subsection{Human Evaluation}\label{appendix: test system}


\begin{figure*}[!h]
  \begin{subfigure}{\linewidth}
    \includegraphics[width=\linewidth]{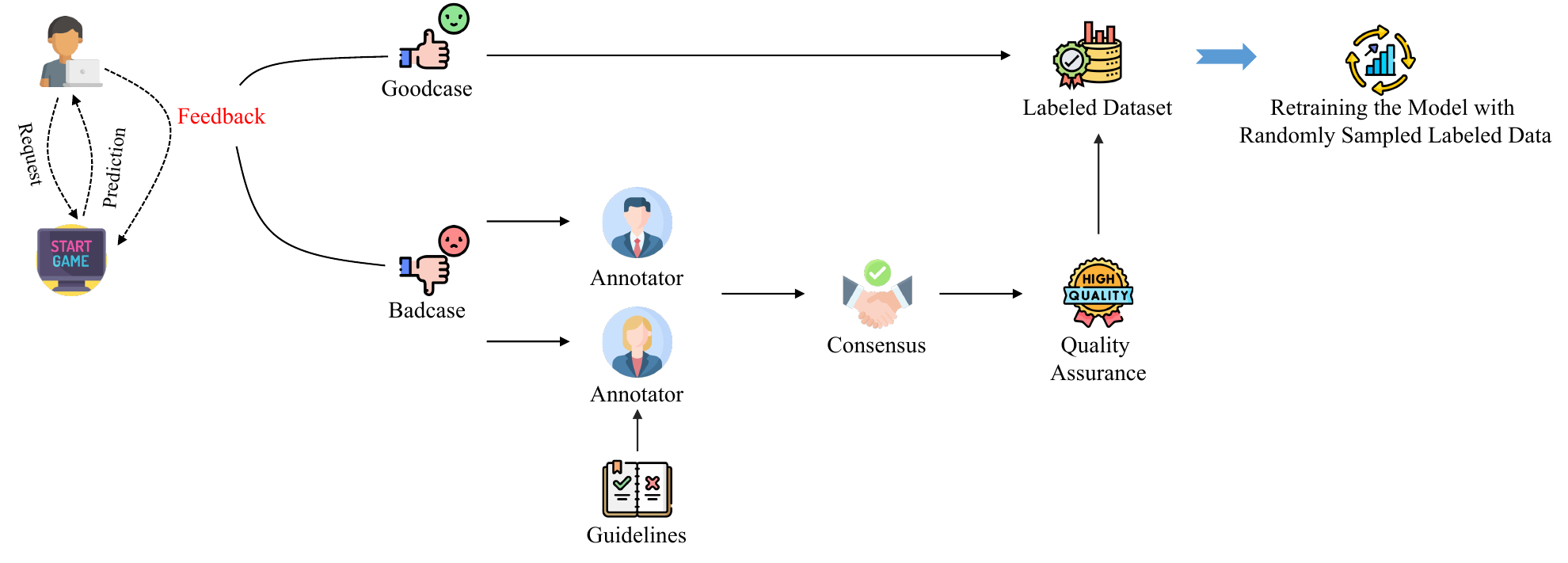}
    \caption{\textbf{The game testing system.} Human players can mark their real preference during gameplay. Good cases will automatically generate a dataset that can be used for model training and evaluation. For bad cases, we will collect dynamic game data and tactical replays used to the annotation system.}
    \label{fig:labeling}
  \end{subfigure}
  \hfill
  \begin{subfigure}{\linewidth}
    \includegraphics[width=\linewidth]{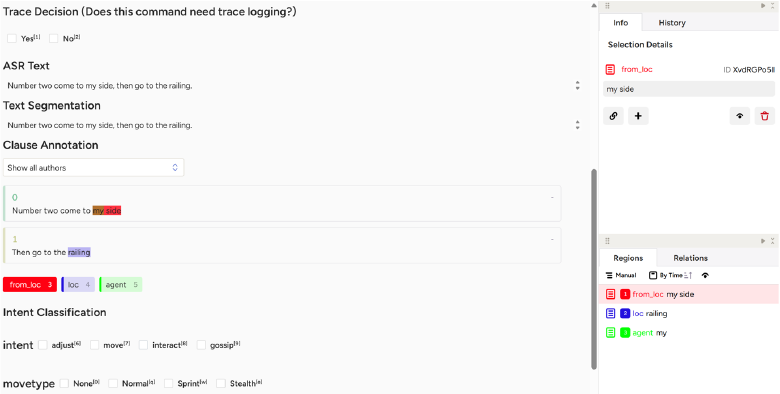}
    \caption{\textbf{The interface of the annotation system.} Annotators can evaluate various modules of \alg, including ASR, intent recognition, and environmental perception.}
    \label{fig:labeling_ui}
  \end{subfigure}
  \caption{Overview of the testing and annotation system}
\end{figure*}

To continuously assess and improve the capabilities of \alg, as well as to collect valuable player interaction data, we have developed a comprehensive human testing and evaluation system illustrated in Fig.~\ref{fig:labeling}. This system integrates real-time feedback from players engaging with the game \textit{Arena Breakout: Infinite}, enabling us to monitor the model’s performance within a realistic gameplay context. During gameplay, testers provide instantaneous feedback on \alg’s outputs. For positively verified cases (“good cases”), the corresponding model outputs are automatically saved as labeled data, which can be directly utilized for further training and evaluation. For negatively evaluated cases (“bad cases”), the same data points are annotated independently by multiple annotators to ensure reliability and reduce bias. Meanwhile, dynamic in-game data and tactical replays related to these cases are recorded and stored in the background to support detailed post-hoc analysis.


As shown in Fig.~\ref{fig:labeling_ui}, the annotation interface allows human annotators to systematically evaluate multiple core components of \alg, including Automatic Speech Recognition (ASR), intent recognition, and environmental perception. To further enhance the quality of the labeled data, we have introduced a detailed tagging mechanism specifically for the entity recognition module. This mechanism helps improve the accuracy and richness of the information related to the game’s dynamic 3D assets. To measure annotation quality and consistency, we employ Inter-Annotator Agreement (IAA) metrics. Only data verified to be of high quality through this agreement process is selected to form the refined labeled dataset used for retraining and optimizing \alg.

Overall, this human evaluation framework provides an effective, scalable, and data-driven method for continuously refining the performance of \alg through real player interactions, detailed annotations, and in-depth replay analysis.

\end{document}